\documentclass[aps,twocolumn,preprintnumbers,bibnotes10pt,superscriptaddress]{revtex4}

\usepackage{xcolor}
\usepackage{amsmath}
\usepackage{graphicx}
\usepackage{ulem}
\usepackage{dcolumn}
\usepackage{epsfig}
\usepackage{bm}
\usepackage{array}
\usepackage[frenchb]{babel}
\usepackage{hyperref}
\hypersetup{
colorlinks=true,
citecolor=blue,
linkcolor=red,
urlcolor=blue, 
bookmarks=true,
pdfmenubar=true
}

\usepackage{graphicx}
\usepackage{amsmath}
\usepackage{bm}           % bold math
\usepackage{bbm}
\usepackage{ulem}

\newcommand{\bra}[1]{\ensuremath{\langle#1|}}
\newcommand{\ket}[1]{\ensuremath{|#1\rangle}}

\newcommand{\vect}[1]{\bm{#1}}

\newcommand{\be}{\begin{equation}}
\newcommand{\ee}{\end{equation}}

\newcommand{\beq}{\begin{eqnarray}}
\newcommand{\eeq}{\end{eqnarray}}

\begin{document}

\title{Robust multipartite entanglement in dirty topological wires}

\author{Luca Pezz\`e}
\email{luca.pezze@ino.cnr.it}
\affiliation{Istituto Nazionale di Ottica, Consiglio Nazionale delle Ricerche (INO-CNR), Largo Enrico Fermi 6, 50125 Firenze, Italy} 
\affiliation{European Laboratory for Nonlinear Spectroscopy (LENS), Via N. Carrara 1, 50019 Sesto Fiorentino, Italy}

\author{Luca Lepori}
\email{luca.lepori@unipr.it}
\affiliation{Dipartimento di Scienze Matematiche, Fisiche e Informatiche, Universit\`a  di Parma, Parco Area delle Scienze, 53/A, I-43124 Parma, Italy}
\affiliation{INFN, Gruppo Collegato di Parma, Parco Area delle Scienze 7/A, 43124, Parma, Italy}

\begin{abstract}
Identifying and characterizing quantum phases in the presence of long-range correlations and/or spatial disorder is, generally, a challenging and relevant task.
Here, we study a generalization of the Kitaev chain with variable-range pairing and different forms of site-dependence of the chemical potential, addressing commensurate and incommensurate modulations as well as Anderson disorder. 
In particular, we analyze multipartite entanglement (ME) in the ground state of dirty topological wires, by studying the scaling of the quantum Fisher information (QFI) with the system size.
For nearest-neighbor pairing, the Heisenberg scaling of the QFI is found in one-to-one correspondence with topological phases hosting Majorana modes. 
For finite-range pairing, we recognize long-range phases by the super-extensive scaling of the QFI and characterize complex lobe-structured phase diagrams. 
 The present work contributes to establish QFI and ME as useful quantities to study intriguing aspects of topological systems, also with long-range pairings and interactions.
More generally, our study reports the robustness of spatial multipartite entanglement  against spatial inhomogeneities. 

\end{abstract}

\maketitle
\date{\today}

\section{Introduction}
\label{intro}

The characterization of quantum phases and quantum phase transitions through entanglement-based approaches is an intriguing problem at the frontier between quantum information~\cite{HorodeckiRMP2009,GuhnePR2009} and many-body physics~\cite{AmicoRMP2008,orus2008,EisertRMP2010,Zeng,LaflorenciePR2016,DeChiaraRPP2018}.
The literature has mostly focused on bipartite entanglement, with witnesses such as the von Neumann entropy~\cite{calabrese2004,LatorreJPA2009,OsbornePRA2002,VidalPRL2003,preskill2006}, the entanglement spectrum~\cite{LiPRL2008,PollmannPRB2010,FidkowskiPRL2010,ThomalePRL2010,lepori2012,leporiLR}, and pairwise entanglement~\cite{OsterlohNATURE2002,LidarPRL2004}, also in the presence of disorder~\cite{chack2010,LevyUNIVERSE2019,liu2016}.
By contrast, multipartite entanglement (ME) has been much less studied~\cite{GuhneNJP2005,HofmannPRB2014}, although it captures a more complex entanglement structure than that identified by bipartite or pairwise entanglement~\cite{orus2014}.
A prominent tool to analyze ME is the quantum Fisher information (QFI)~\cite{PezzePRL2009,HyllusPRA2012,TothPRA2012}, which is also central to quantum metrology~\cite{PezzeRMP,TothJPA2014}.
The QFI has been investigated in symmetry-protected and genuinely topological systems~\cite{pezze2017,ZhangPRL2018,ZhangARXIV,LambertPRL2020,YangARXIV,MeraSCIPOST2022}, as well as in spin~\cite{hauke2016,MaPRA2009,LiuJPA2013,pezze2018,MathewPRR2020,LaurellPRL2021} and lattice~\cite{lucchesi2019,daley2020} setups, and has been used to characterize interesting many-body phenomena such as quantum criticality~\cite{FrerotPRL2018,gabbrielli2018,FrerotNATCOMM2019}, quantum chaos~\cite{LerosePRA2020,LiPRA2021,ShiSCIPOST2025}, quantum quenches~\cite{StrobelSCIENCE2014,PappalardiJSM2017}, scrambling~\cite{GarttnerPRL2018}, and thermalization~\cite{GietkaPRB2019,BrenesPRL2020}.

An essential property of the above-mentioned topological phases of matter~\cite{topins2009,hasankane} is their robustness against local perturbations~\cite{PotterPRL2010,hastings2010}.
For instance, noninteracting symmetry-protected topological phases~\cite{ChiuRMP2016} are characterized by the presence of robust gapless boundary states.
The number of such protected states is proportional to the topological invariant characterizing the system and thus plays the role of a global order parameter.
This invariant changes only across a quantum phase transition, through the closing and reopening of the bulk energy gap, and is therefore immune to small but finite local perturbations.
One of the most important examples of a symmetry-protected topological system is the Kitaev chain~\cite{KitaevPU2001,AliceaRPP2012}, a celebrated tight-binding model of one-dimensional spinless fermions with $p$-wave superconductivity.
This prototype hosts nonlocal Majorana modes localized at the edges of an open chain, which exhibit non-Abelian exchange statistics under braiding~\cite{IvanovPRL2001,AliceaNATPHYS2011}.
The intrinsic robustness of topological properties against local perturbations holds the promise of enabling resilient quantum information processing~\cite{KitaevAP2003,SarmaNPJQI2015,NayakRMP2008}, including topologically protected qubits and gate operations~\cite{PachosBOOK,sekania2017}.

So far, the robustness of topological phases in the Kitaev chain has been mainly assessed by studying the fate of Majorana modes under sufficiently strong local perturbations, such as spatial inhomogeneities~\cite{DeGottardiNJP2011,DeGottardiPRL2013,DeGottardiPRB2013,LangPRL2012,TezukaPRB2012,CaiPRL2013,CaiJP2014,hu2015}, as well as in the presence of interactions~\cite{LobosPRA2012,GergsPRB2016,McGinleyPRB2017} or long-range pairing~\cite{CaiJP2017,MishraJPA2020,frax2021}.
However, beyond edge states, the Kitaev model also displays nontrivial quantum-entanglement properties~\cite{VidalPRL2003,calabrese2004,HastingsJSM2007,vodola2014,leporiLR,pezze2017,AresPRA2018}.
Is this entanglement robust against local perturbations? 
And, if so, is it as robust as the Majorana modes?

In this work, motivated by these questions, we investigate the stability of ME against spatial inhomogeneities in a generalization of the Kitaev chain with superconducting pairing decaying with distance as a power law~\cite{vodola2014,vodola2016}.
We consider commensurate and incommensurate quasiperiodic modulations of the chemical potential, as well as uncorrelated on-site disorder drawn from an Anderson distribution.
These perturbations preserve the symmetries of the Hamiltonian, namely charge conjugation and ${\bf Z}_2$ fermion-number parity.
We characterize various short-range (SR) and long-range (LR) phases through the different scaling of the QFI with system size.
We find that the superextensive scaling of the QFI is robust against inhomogeneities and persists up to modulation strengths large enough to induce a quantum phase transition.
In particular, inhomogeneities induce complex structures, such as reentrances and lobes, in the phase diagrams, and in some cases even extend the topological phases.
Furthermore, tuning the pairing range can induce a transition from SR to LR phases without closing the mass gap: this SR-to-LR transition is captured by a change in the scaling of the QFI, also in the presence of spatial inhomogeneities.
Our results suggest that the QFI is a useful tool for characterizing even more complex topological systems, especially when analyzed jointly with other theoretical tools.
This possibility may be even more relevant for long-range systems, where topological invariants are generally difficult to define, especially in the absence of translational invariance.
More broadly, our study provides, to the best of our knowledge, the first evidence of the robustness of ME against spatial inhomogeneities.

\section{Model and entanglement witnesses}
\label{model}

We study a generalization of the Kitaev chain~\cite{KitaevPU2001} with an inhomogeneous (site-dependent) chemical potential $\mu_l$ and algebraically decaying superconducting pairing~\cite{vodola2014,vodola2016}, characterized by the exponent $\alpha$:
\begin{align}
 \label{H}
\begin{split}
\hat{H} = - \frac{J}{2} \sum_{l=1}^{L} \big(\hat{a}_l^\dag \hat{a}_{l+1} + \hat{a}_{l+1}^\dag \hat{a}_l \big) - \sum_{l=1}^L \mu_l  \hat{a}_l^\dag \hat{a}_l \, + \\
+ \, \frac{\Delta}{2}  \sum_{m=1}^L \sum_{l=1}^{L-1} d_l^{-\alpha} \, \big( \hat{a}_m \hat{a}_{m+l}  + \hat{a}_{m+l}^\dag \hat{a}_m^\dag \big) \, .
\end{split}
\end{align}
Here, $\hat{a}_l^\dag$ ($\hat{a}_l$) is the creation (annihilation) operator of a spinless fermion at site $l$, $L$ is the number of lattice sites, $J>0$ is the hopping amplitude, $\Delta$ is the pairing strength and $d_l = l$ for an open chain, while $d_l = \mathrm{min} (l, L-l)$ for a closed chain with periodic boundary conditions, in which case the site indices are understood modulo $L$.
In the limit $\alpha \to \infty$, one has $d_l^{-\alpha} \to \delta_{l,1}$, where $\delta_{l,m}$ is the Kronecker delta, thus recovering nearest-neighbor pairing.
Following Ref.~\cite{Lieb1961}, the Hamiltonian in Eq.~(\ref{H}) can be rewritten in quadratic form, and the model can be diagonalized exactly for arbitrary values of $\alpha$ and $\mu_l$; see footnote~\cite{nota1} for details.

The model in Eq.~(\ref{H}) has been extensively studied in the literature, especially for a uniform chemical potential, $\mu_l=\mu$, and in the limit $\alpha \to \infty$.  
A quantum phase transition at $|\mu|/J = 1$ separates a topological phase (for $|\mu|/J < 1$) from a topologically trivial phase (for $|\mu|/J > 1$)~\cite{KitaevPU2001}.  
For open boundary conditions, the topological phase is characterized by Majorana edge modes with anyonic statistics~\cite{KitaevPU2001,stern2008}, related to a ${\bf Z}_2$ degeneracy of the ground state in the thermodynamic limit.  
The system belongs to class $D$ in the classification of topological insulators and superconductors and is characterized by particle-hole symmetry $C$~\cite{topins2009,hasankane}.  
The topological phase and the associated edge modes are topologically protected.  
Specifically, they are stable against perturbations, such as chemical-potential inhomogeneities, provided that these are weak compared to the mass gap and do not alter the symmetry, and hence the symmetry-protected topological nature, of the quantum state~\cite{PotterPRL2010}.  
The main effect of spatial inhomogeneities is to reduce the spectral gap to the lowest-energy excited state.  
For sufficiently strong perturbations, however, the excitation gap eventually closes and the Majorana modes disappear.  
The inhomogeneity-induced quantum phase transition associated with the fate of the Majorana modes has been studied in Refs.~\cite{DeGottardiNJP2011,DeGottardiPRL2013,DeGottardiPRB2013,LangPRL2012,TezukaPRB2012,CaiPRL2013,CaiJP2014,hu2015} for nearest-neighbor coupling and different distributions of $\mu_l$.

Variable-range extensions ($\alpha<\infty$) of the Kitaev chain have also received considerable attention, especially in the case of a uniform chemical potential, $\mu_l=\mu$~\cite{vodola2014,vodola2016,lepori2016eff,leporiLR,pezze2017}.
In this case, and for $1<\alpha<\infty$, the system is known to be topologically equivalent to the SR Kitaev chain~\cite{leporiLR}: the topological Majorana phase for $|\mu|/J<1$ persists down to $\alpha=1$.
For $\alpha<1$, the situation changes qualitatively.
The corresponding phase diagram, for a homogeneous chemical potential, has been studied in Refs.~\cite{vodola2014,pezze2017}.
Purely LR insulating phases, namely phases not included in the classification of SR topological insulators~\cite{altland1997}, emerge.
A quantum phase transition is found at $\mu/J=1$.
The phase for $\mu/J<1$ hosts massive edge modes, originating from the hybridization of Majorana modes~\cite{vodola2016,pachos2017} and also referred to as Dirac modes, even in the thermodynamic limit; this phase is associated with a LR topology~\cite{leporiLR}.
For $\mu/J>1$, a LR phase with no edge modes is present.
Interestingly, when $\alpha$ is varied across $\alpha=1$, the transitions between SR and LR phases occur without closing the mass gap.
We recall that the above behavior holds for a uniform chemical potential.
Less is known about the interplay between long-range pairing and chemical-potential inhomogeneities, and about the robustness of LR phases~\cite{CaiJP2017,MishraJPA2020,frax2021}.

\subsection*{Quantum Fisher information}
\label{QFI}

The QFI can be defined as the susceptibility of the quantum (Uhlmann) fidelity $\mathcal{F}(\theta,\mathrm{d}\theta)$~\cite{Uhlmann1976,Jozsa1994} between a quantum state depending on a parameter $\theta$ and its infinitesimally displaced neighbor at $\theta+\mathrm{d}\theta$~\cite{BraunsteinPRL1994,PezzePNAS2016,Sidhu2020}.
The QFI is 
\be
F_Q[|\psi(\theta)\rangle]
=
8\lim_{\mathrm{d}\theta\to 0}
\frac{1-\mathcal{F}(\theta,\mathrm{d}\theta)}{(\mathrm{d}\theta)^2} \, ,
\ee
where 
\begin{equation}
\label{qidef0}
\mathcal{F}(\theta,\mathrm{d}\theta)
=
\big|\langle\psi(\theta)\mid\psi(\theta+\mathrm{d}\theta)\rangle\big|
\end{equation}
is the fidelity between pure states, as relevant for this work. 
A convenient choice of parametric transformation is the unitary encoding $|\psi(\theta)\rangle = e^{-i\hat J\theta}|\psi\rangle$.
Here, $\hat J=\sum_{l=1}^L \hat j_l$ is a collective operator, where $l=1,\dots,L$ labels the $L$ subsystems and $\hat j_l$ is a Hermitian operator acting on the local Hilbert space $\mathcal H_l$ ($\mathcal H=\bigotimes_{l=1}^L \mathcal H_l$ being the total Hilbert space of the system).
This choice leads to
\begin{equation}
\label{qidef}
F_Q[|\psi\rangle,\hat J]
=
4\sum_{l,m}
\big(
\langle\psi|\hat j_l\hat j_m|\psi\rangle
-
\langle\psi|\hat j_l|\psi\rangle
\langle\psi|\hat j_m|\psi\rangle
\big) \, ,
\end{equation}
namely, the QFI is given by the sum of the connected two-point correlations of the local generators~\cite{pezze2014}.
The scaling of the QFI with system size is therefore closely related to the decay of connected correlations.
Notice, however, that due to finite-size effects, the former is often much easier to investigate directly than the latter.

The QFI in Eq.~\eqref{qidef} is an efficient witness of ME~\cite{PezzePRL2009,HyllusPRA2012,TothPRA2012,notaME}:
for any $k$-producible state, it satisfies the inequality
\begin{equation}
\frac{F_Q[\ket{\psi},\hat J]}{(\Delta \hat j)^2} \le s k^2 + r^2\leq kL \, ,
\label{boundQFI}
\end{equation}
where $L = s k + r$, $0 \le r < k$, $(\Delta \hat j)^2 \equiv (j_{\max}-j_{\min})^2$ is the square of the difference between the maximum and minimum eigenvalues of the local operator $\hat j_l$, and we assume that all operators $\hat j_1,\dots,\hat j_L$ have the same bounded spectrum.
Therefore, a violation of Eq.~\eqref{boundQFI} signals an entanglement depth greater than $k$, namely at least $(k+1)$-partite entanglement among the $L$ subsystems $\mathcal{H}_l$~\cite{HyllusPRA2012,TothPRA2012}.
More explicitly, separable states, such as pure product states $\ket{\psi}_{\rm sep}=\bigotimes_{l=1}^L \ket{\psi_l}$, satisfy~\cite{PezzePRL2009}
\be
\frac{F_Q[\ket{\psi}_{\rm sep},\hat J]}{(\Delta \hat j)^2} \le L \, .
\ee
In other words, for separable states the QFI scales, at best, extensively with system size.
A superextensive scaling, namely
\[
\frac{F_Q[\ket{\psi},\hat J]}{(\Delta \hat j)^2}\sim L^\beta
\qquad \text{with } \beta>1 \, ,
\]
is only possible if $\ket{\psi}$ is multipartite entangled, with entanglement depth at least $k\sim L^{\beta-1}$.
Notice, however, that ME is not necessarily associated with a 
superextensive scaling of the QFI.
Finally,
\be
\frac{F_Q[\ket{\psi},\hat J]}{(\Delta \hat j)^2}\sim L^2
\ee
is the fastest possible scaling of the QFI for the collective local operators considered here: it is known as Heisenberg scaling~\cite{PezzePRL2009} and is associated with $k\sim L$.
Importantly, the entanglement witnessed by the QFI can be exploited for metrology and sensing in terms of the local operator; see, e.g., Ref.~\cite{pezze2014}.
In particular, the QFI bounds the corresponding achievable parameter-estimation sensitivity.
Finally, the direct quantitative link between entanglement and correlations is, in general, much less explicit than that provided by the QFI.

In the following, we consider the state $\ket{\psi}$ being the ground state, $\ket{\psi_{\rm gs}}$, of Eq.~(\ref{H}).
As the operator $\hat J$, we consider the family of collective pseudo-spin operators
\be \label{Jops}
\hat{J}_{x,y}(\vect{s}) = \sum_{l=1}^L s_l \frac{\hat{\sigma}_{x,y}^{(l)}}{2} \, ,
\ee
where $\vect{s}=\{s_1,\dots,s_L\}$ and $s_l=\pm 1$ are local sign coefficients.
This choice corresponds to the local generators $\hat j_l = s_l \frac{\hat{\sigma}_{x,y}^{(l)}}{2}$, for which $(\Delta \hat j)^2=1$.
The Pauli operators can be expressed in terms of spinless fermionic operators through the Jordan-Wigner transformation,
\begin{subequations}
\begin{align}
& \hat{\sigma}_{x}^{(l)}=
\hat{a}_l^\dag e^{i \pi \sum_{m=1}^{l-1} \hat{a}_m^\dag \hat{a}_m}
+
e^{-i \pi \sum_{m=1}^{l-1} \hat{a}_m^\dag \hat{a}_m} \hat{a}_l \, ,\\
& \hat{\sigma}_{y}^{(l)}=
-i \hat{a}_l^\dag e^{i \pi \sum_{m=1}^{l-1} \hat{a}_m^\dag \hat{a}_m}
+
i e^{-i \pi \sum_{m=1}^{l-1} \hat{a}_m^\dag \hat{a}_m} \hat{a}_l \, ,
\end{align}
\end{subequations}
and are therefore highly nonlocal in the fermionic lattice modes.
Taking into account that
$\bra{\psi_{\rm gs}} \big(\hat{\sigma}_{x,y}^{(l)}\big)^2 \ket{\psi_{\rm gs}} = 1$ and  $\bra{\psi_{\rm gs}} \hat{\sigma}_{x,y}^{(l)} \ket{\psi_{\rm gs}} = 0$
for all $l$---the latter as a consequence of fermion-number parity conservation, see, e.g., Ref.~\cite{mussardo}---we obtain
\be \label{Fxy}
F_Q[\ket{\psi_{\rm gs}}, \hat{J}_{x,y}(\vect{s})]
=
\sum_{l,m=1}^{L} s_l s_m \,\rho_{x,y}^{(l,m)} .
\ee
In the translationally invariant case, this expression reduces to
\be \label{Fxytrans}
F_Q[\ket{\psi_{\rm gs}}, \hat{J}_{x,y}(\vect{s})]
=
L + L \sum_{l=1}^{L-1} s_l s_1 \,\rho_{x,y}^{(l,1)} .
\ee
Following Ref.~\cite{Lieb1961}, the correlation functions can be calculated as
\begin{equation}
\rho_{x}^{(l,m)} = {\rm det}
\left( \begin{array}{cccc}
G_{l,l+1} & G_{l,l+2} & \dots & G_{l,m} \\
\vdots & \vdots &  & \vdots \\
G_{m-1,l+1} & G_{m-1,l+2} & \dots & G_{m-1,m}
\end{array} \right) ,
\end{equation}
and
\begin{equation}
\rho_{y}^{(l,m)} = {\rm det}
\left( \begin{array}{cccc}
G_{l+1,l} & G_{l+1,l+1} & \dots & G_{l+1,m-1} \\
\vdots & \vdots &  & \vdots \\
G_{m,l} & G_{m,l+1} & \dots & G_{m,m-1}
\end{array} \right) ,
\end{equation}
where $\vect{G} = - \vect{\Psi}^{\rm T}\vect{\Phi}$; see footnote~\cite{nota1}.
In Appendix~A we show that Eq.~(\ref{Fxy}) witnesses $k$-partite entanglement, for pure as well as mixed states, through the same bound as in Eq.~\eqref{boundQFI}.
We are mainly interested in the power-law scaling of $F_Q$ with the system size $L$.
We therefore define the QFI scaling exponent
\be \label{defbetaxy}
\beta_{x,y}(\vect{s})=
\frac{d \log F_Q[\ket{\psi_{\rm gs}}, \hat{J}_{x,y}(\vect{s})]}{d \log L} \, .
\ee

Let us recall some previous results concerning the scaling of the QFI for the uniform chain, $\mu_l=\mu$, and $\Delta \neq 0$~\cite{pezze2017}. 
For SR coupling, $\alpha > 1$, it has been shown that in the topological phase, $|\mu|/J < 1$, the QFI is maximized by the operator $\hat{J}_x = \sum_{l=1}^L \hat{\sigma}_x^{(l)}/2$ [namely, by choosing the uniform $\vect{s}_{\rm unif}=\{1,1,\dots,1\}$ in Eq.~(\ref{Jops})], and the corresponding scaling exponent in Eq.~\eqref{defbetaxy} is $\beta_x(\vect{s}_{\rm unif})=2$.
The operator $\hat{J}_x$ is the order parameter of the quantum Ising chain in a transverse field, which corresponds, via the Jordan-Wigner transformation, to the standard ($\alpha \to \infty$) Kitaev chain~\cite{thomale2014,mussardo}.
Along the critical lines $\mu/J = \pm 1$, the QFI scales with exponent $\beta_x(\vect{s}_{\rm unif})=3/2$, while in the trivial phase, $|\mu|/J > 1$, the scaling is $\beta_{x,y}(\vect{s}_{\rm unif})=1$.
For LR coupling, namely $0 \leq \alpha < 1$, and $\mu/J<1$, the QFI is still maximized by the operator $\hat{J}_x = \sum_{l=1}^L \hat{\sigma}_x^{(l)}/2$ and has scaling exponent $\beta_x(\vect{s}_{\rm unif})=7/4$.
By contrast, for $\mu/J>1$, the QFI is maximized by the operator $\hat{J}_y(\vect{s}_{\rm stag})$ with staggered $\vect{s}_{\rm stag}=\{-1,1,-1,\dots\}$, yielding $\beta_y(\vect{s}_{\rm stag})=7/4$.
Along the critical line $\mu/J = 1$, one again finds the scaling exponent $\beta_x(\vect{s}_{\rm unif}) = \beta_y(\vect{s}_{\rm stag}) = 3/2$.
A scaling $\beta_{x,y}(\vect{s})>1$ [$\beta_{x,y}(\vect{s})=1$] is associated with algebraically [exponentially] decaying two-point correlations of the $\hat{\sigma}_{x,y}$ operator~\cite{pezze2017}. 

In the non-uniform case discussed below -- especially for LR coupling -- two-point correlation functions become more involved.
For instance, in certain cases, $\rho_{x,y}^{(l,1)}$ is characterized, when varying $l$, by clusters of positive values followed by clusters of negative ones, rather than the alternating staggered distribution observed in the uniform case, as mentioned above.
For this reason, we compute 
\be \label{Fxyabs}
\tilde{F}_Q[\ket{\psi_{\rm gs}}, \hat{J}_{x,y}]
=
\sum_{l,m=1}^{L} |\rho_{x,y}^{(l,m)}| \, ,
\ee
which involves the sum of the moduli of the connected correlations of the usual spin operators.
In the translationally invariant case, this becomes
\be 
\tilde{F}_Q[\ket{\psi_{\rm gs}}, \hat{J}_{x,y}]
=
L + L \sum_{l=1}^{L-1} |\rho_{x,y}^{(l,1)}| \, .
\ee
These quantities generalize Eqs.~(\ref{Fxy}) and (\ref{Fxytrans}), respectively.
In Appendix~A we show that Eq.~(\ref{Fxyabs}) is a $k$-partite entanglement witness for pure states.  
In particular, we study the scaling exponent
\be \label{defbetabsxy}
\beta_{x,y}=
\frac{d \log \tilde{F}_Q[\ket{\psi_{\rm gs}}, \hat{J}_{x,y}]}{d \log L} \, .
\ee
We generally find that, for $\alpha>1$ the largest scaling exponent is obtained for $F_Q[\ket{\psi_{\rm gs}}, \hat{J}_x(\vect{s}_{\rm unif})]$. 
We denote the region of parameters having $\beta_x = \beta_x(\vect{s}_{\rm unif})>1$ as xSR phase.
We also typically find that $\beta_y=1$ in that regime.
For $0\leq \alpha<1$, the regions of parameters for which $\beta_{x,y}>1$ are indicated as xLR and yLR, respectively.
We also indicate 
\be \label{eqbeta}
\beta = \max(\beta_x,\beta_y) \, .
\ee

\subsection*{Topological invariant in the $\alpha \to \infty$ limit}

In the limit $\alpha \to \infty$, we compare the predictions of the QFI with the topological phases identified by a ${Z}_2$ topological index $\nu$.
Assuming (anti-)periodic boundary conditions, $\nu$ can be computed in various equivalent ways~\cite{bernevigbook}.
For instance, one common approach is through the Berry phase~\cite{berry1984},
\beq
\nu = \frac{i}{\pi} \int_{\rm BZ} \mathrm{d}k \, \langle u_k | \partial_k u_k \rangle \, ,
\label{berryphase}
\eeq
accumulated as $k$ varies across the first Brillouin zone.
Here, $|u_k\rangle$ is the positive-energy eigenvector of $\hat{H}(k)$---or, equivalently, the negative-energy one---where $\hat{H}(k)$ is the Fourier transform of Eq.~\eqref{H}.
Assuming instead open boundary conditions, $\nu$ can be obtained from the transfer matrix equation~\cite{DeGottardiNJP2011, DeGottardiPRL2013, DeGottardiPRB2013}
\be
\begin{pmatrix}
\psi_{j+1} \\
\psi_{j}
\end{pmatrix} = D_j 
\begin{pmatrix}
\psi_{j} \\
\psi_{j-1}
\end{pmatrix},  \quad \quad
D_j = \begin{pmatrix}
\frac{\mu_j}{\Delta + t} \quad \frac{\Delta - t}{\Delta + t} \\
1 \quad 0
\end{pmatrix} \, , 
\ee
where $\psi_{i}$ is the eigenfunction of the closest positive-energy eigenfunction above zero energy. 
In the topological phase, this solution is associated with Majorana edge modes.
For a system of $L$  lattice sites, one considers the matrix $\mathcal{D}  = \prod_{i=1}^{L} D_i$.
Let $\lambda_1$ and $\lambda_2$ be the two eigenvalues of $\mathcal{D}$, ordered such that $|\lambda_1|<|\lambda_2|$.
One can then define the topological invariant
\be
\nu = \frac{1-\mathrm{sgn}\!\left(\ln |\lambda_2|\right)}{2} \, .
\label{calcnu}
\ee
This quantity takes the value $\nu=0$ in the trivial superconducting phase and $\nu=1$ in the topological phase.
The index $\nu$ is related to the ${\bf Z}_2$ parity structure of the ground state and, by bulk-boundary correspondence, signals the presence or absence of Majorana edge modes.

We note that, at least in the homogeneous case $\mu_j=\mu$, the topological invariant in Eqs.~\eqref{berryphase} and \eqref{calcnu} can also be expressed in terms of $\zeta = \mathrm{Pf}[\hat{M}]/|\mathrm{Pf}[\hat{M}]|$~\cite{KitaevPU2001,prodan2014}, where $\mathrm{Pf}[\hat M]$ is the Pfaffian of the Hamiltonian matrix recast in a skew-symmetric form $\hat M$ by adopting a Majorana site-operator basis~\cite{bernevigbook}.
As the Hamiltonian parameters are varied, the Pfaffian can change sign when the bulk gap closes.
Therefore, $\zeta$ can also be used as a ${\bf Z}_2$ topological invariant, as required by the class-$D$ symmetry of the Hamiltonian in Eq.~\eqref{H}~\cite{topins2009}, taking the values $\pm 1$.
For open boundary conditions, $\zeta$ is likewise related to the ${\bf Z}_2$ fermion-parity structure of the ground state, similarly to the index $\nu$.
Operationally, $\zeta$ can be evaluated through a closed formula for even-dimensional matrices, as discussed in Ref.~\cite{wimmer2012prova}.
However, when perturbations explicitly breaking translational invariance are introduced, the simple momentum-space Pfaffian expression is no longer directly applicable as a topological index.
This relevant situation will be further discussed in Appendix B, also for finite $\alpha$.
Finally, other equivalent expressions for $\nu$ are known in the translationally invariant regime, both in the thermodynamic limit~\cite{obuse2013} and at finite size~\cite{suzuki2005}.

%%%%%%%%%%%%%%%%%%%%%%%%%%%%%%%%%%
%% FIGURE 1
%%%%%%%%%%%%%%%%%%%%%%%%%%%%%%%%%%
\begin{figure*}[t!]
\begin{center}
\includegraphics[width=1\textwidth]{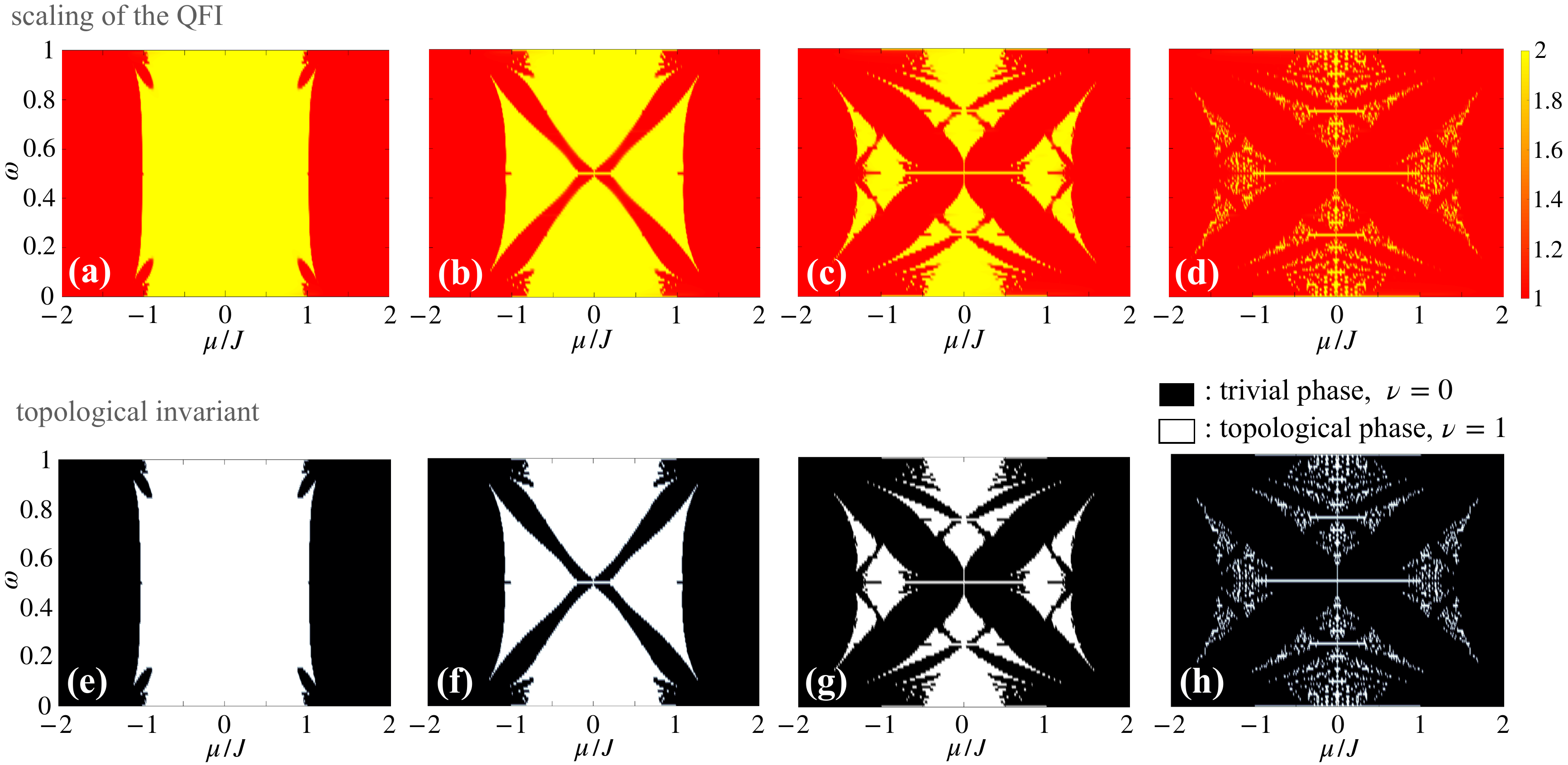}
\end{center}
\caption{
Phase diagram of the model in Eq.~(\ref{H}) with nearest neighbor coupling ($\alpha=\infty$), as a function of the commensurate modulation frequency $\omega = p/q$.
Upper panels: scaling exponent $\beta_x$ of the QFI (colorscale).
Lower panels: topological invariant $\nu$, defined in Eq.~\eqref{calcnu}: white regions correspond to topological phases with $\nu=1$, black regions to trivial phases with $\nu=0$.
Panels (a,e), (b,f), (c,g), and (d,h) correspond respectively to $V/J=0.25$, $0.5$, $1$, and $1.25$.
Here, $\Delta/J=0.25$ and the phase $\phi$ in Eq.~(\ref{muj_AA}) is set to $\phi=0$.
The scaling $\beta$ is extracted from the calculation of the QFI between $L=200$ and $L=400$ with antiperiodic boundary conditions.
The invariant $\nu$ is computed for $L=400$ with open boundary conditions.
Here $\omega$ is changed according to $p=0,1,2, ..., 100$ and $q=100$.
}
\label{Fig1}
\end{figure*}
%%%%%%%%%%%%%%%%%%%%%%%%%%%%%%%%%%
%%%%%%%%%%%%%%%%%%%%%%%%%%%%%%%%%%
%%%%%%%%%%%%%%%%%%%%%%%%%%%%%%%%%%

\section{Commensurate quasiperiodic modulation (Harper potential)}
\label{commpot}

We analyze here the following site dependence of the chemical potential:
\be \label{muj_AA}
\mu_l = \mu + V \sin \big(2\pi l \omega + \phi \big) \, ,
\ee
where $\mu$ is a constant offset, $\omega = p/q$ is the commensurate modulation frequency, with $p$ and $q$ relatively prime integers, $V$ quantifies the strength of the inhomogeneity, and $\phi \in [0,2\pi)$ is a phase.
The salient features of the topological phase diagrams discussed in this section depend only weakly, at the qualitative level, on $\phi$.

For $\Delta = 0$, the Hamiltonian in Eq.~\eqref{H}, with the Harper potential in Eq.~\eqref{muj_AA} and rational $\omega = p/q$~\cite{Harper1955}, maps exactly onto a square-lattice tight-binding model with a uniform magnetic flux $p/q$ per plaquette (in units of the flux quantum $h/e$, where $e$ is the electric charge)~\cite{hofstadter1976}.
In this mapping, $V$ plays the role of the hopping amplitude along a second cyclic coordinate $m$, orthogonal to the chain direction, while $\phi$ is the corresponding transverse quasi-momentum. 
Depending on the ratio $p/q$ and on the filling, the present setup hosts, besides various metallic phases, several insulating phases, possibly also supporting edge excitations~\cite{hofstadter1976,LangPRL2012}. 
These phases appear in a characteristic fractal butterfly-like structure in the plane of the single-particle energies as functions of the magnetic flux. 
For $\Delta \neq 0$, the same mapping as in Ref.~\cite{hofstadter1976} still holds, with the addition of a pairing term along the chain direction.

\subsection*{Limit $\alpha \to \infty$}

The upper panels of Fig.~\ref{Fig1} show the scaling exponent of the QFI as a function of $\mu/J$ and $\omega$, for different values of $V/J$.
The QFI is maximized by the operator $\hat{J}_x$, which is also a nonlocal order parameter in the case $V/J=0$~\cite{pezze2017}.
In Fig.~\ref{Fig1}(a)-(d), we show the corresponding scaling exponent $\beta_x$.
The lower panels report the corresponding topological index $\nu$, calculated as in Eq.~\eqref{calcnu}.
We find a remarkable one-to-one correspondence between nontrivial topology, $\nu=1$, and Heisenberg scaling, $\beta_x = 2$.
Conversely, the trivial phase is characterized by an extensive scaling, $\beta_x =\beta_y =1$.
Recalling that $\phi=0$ is assumed here, for $\omega=0$, $1/2$ and $1$, the sinusoidal modulations in Eq.~(\ref{muj_AA}) vanish and we recover the homogeneous chemical potential case $\mu_l=\mu$.
Accordingly, the topological phase is found for $\vert \mu \vert/J <1$~\cite{KitaevPU2001, pezze2017}, irrespective of $V$, as seen in Fig.~\ref{Fig1}.
The robustness of Majorana edge states against the modulation strength $V$, for specific values of $\phi$ and $\omega$, was also noted in Ref.~\cite{LangPRB2012}.

In the limit $V/J \to 0$ [e.g. $V/J=0.25$ in panels (a) and (e)], we recover the nontrivial topological phase for $|\mu|/J < 1$ and any value of $\omega$.
Increasing $V/J$, we find a typical Hofstadter-butterfly-like phase diagram.
This characteristic structure was already noted in Ref.~\cite{DeGottardiPRL2013} in the study of the topological invariant $\nu$.
For large values of $V/J$ [e.g. $V/J=1.25$ in panels (d) and (h)], the width of the topological phases becomes progressively smaller and eventually vanishes for $V/J \gg 1$.

Note that the rich butterfly structure shown in Fig.~\ref{Fig1} disappears for $\Delta/J \geq 1$ (in Fig.~\ref{Fig1} we have $\Delta/J = 0.25$).
In this case, one obtains a single connected topological region for $|\mu|/J \lesssim 1$, irrespective of $\omega$, for all values of $V/J \lesssim 1$.
By contrast, for $\Delta/J \geq 1$ and $V/J \gtrsim 1$, the phase diagram acquires a dispersed multifractal structure.

%%%%%%%%%%%%%%%%%%%%%%%%%%%%%%%%%%
%% FIGURE 2
%%%%%%%%%%%%%%%%%%%%%%%%%%%%%%%%%%
\begin{figure}[t!]
\begin{center}
\includegraphics[width=\columnwidth]{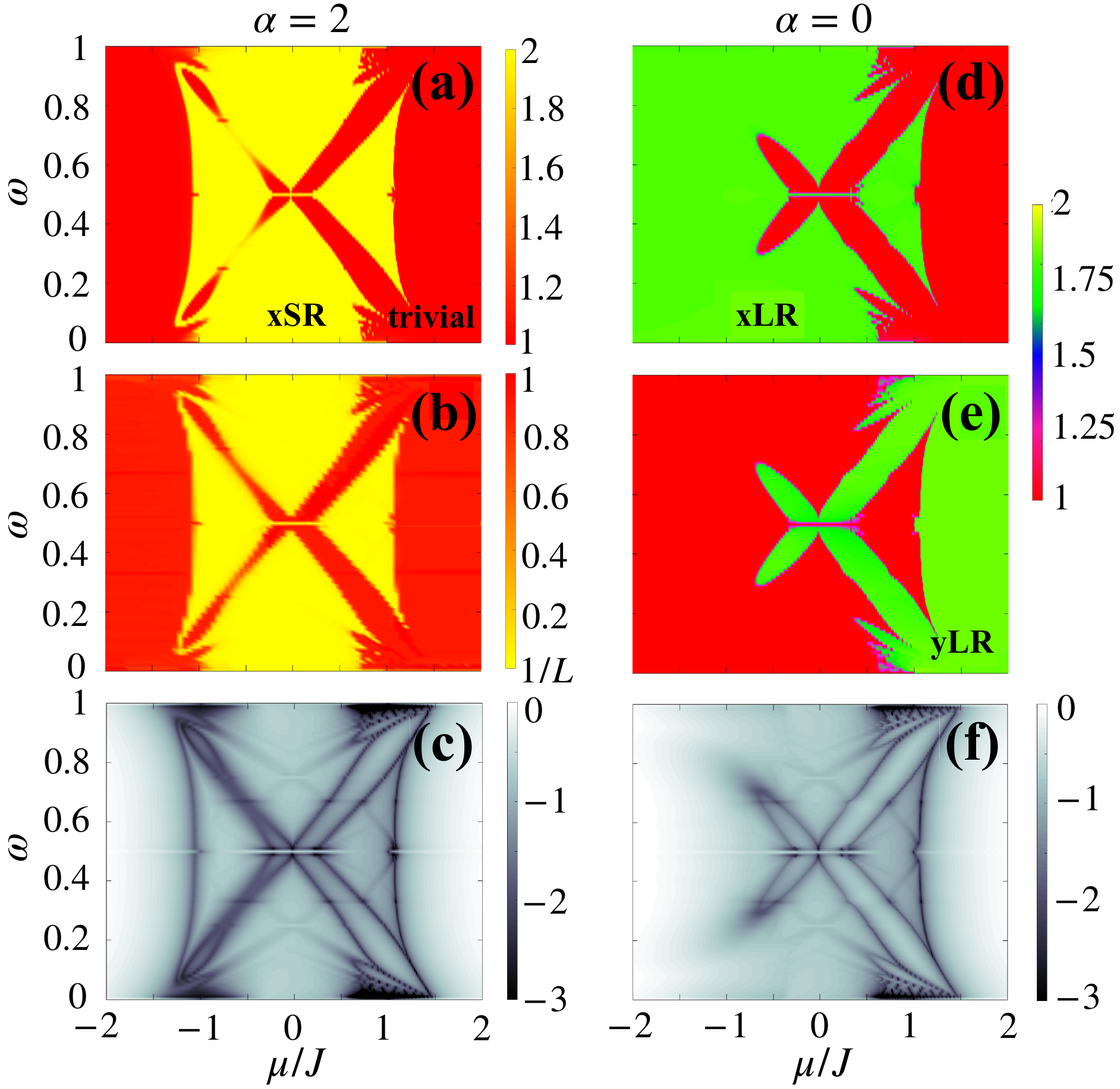}
\end{center}
\caption{
Phase diagram of the model in Eq.~(\ref{H}) with power-law couplings of finite decay exponent~($\alpha<\infty$) and commensurate modulation frequency, shown as a function of $\mu/J$ and $\omega=p/q$.
Panels~(a), (b), and (c) show the scaling exponent $\beta_x$ of the QFI, the ELW $\delta \ell/L$ [see Eq.~(\ref{Eq.Loc}) and the main text], and the logarithm of the mass gap $\log_{10}(\delta E/J)$, respectively, for the SR case $\alpha=2$.
Panels~(d) and (e) show $\beta_x$ and $\beta_y$, respectively, for $\alpha=0$.
Panel~(f) shows the corresponding mass gap, $\log_{10}(\delta E/J)$.
Here, $V/J=0.5$, $\Delta/J=0.25$, $\phi=0$, and the QFI scaling exponent is extracted from the system-size dependence between $L=200$ and $L=400$.
The mass gap and the ELW are calculated at $L=200$.
The QFI and the mass gap are computed for a closed chain with anti-periodic boundary conditions, whereas the ELW is computed for an open chain.
The quantity $\delta E/J$ is plotted with a lower cutoff of $10^{-3}$.
}
\label{Fig2}
\end{figure}
%%%%%%%%%%%%%%%%%%%%%%%%%%%%%%%%%%
%%%%%%%%%%%%%%%%%%%%%%%%%%%%%%%%%%
%%%%%%%%%%%%%%%%%%%%%%%%%%%%%%%%%%

\subsection*{Finite-$\alpha$ regime: $\alpha>1$}

Figure~\ref{Fig2} investigates how the butterfly-like phase diagram of Fig.~\ref{Fig1} changes for finite $\alpha$.
Figure~\ref{Fig2}(a) shows the scaling exponent $\beta_x$ as a function of $\mu/J$ and $\omega$ in the SR case $\alpha=2$.
Here, we focus on the parameter set $\Delta/J=0.25$ and $V/J=0.5$, which, in the limit $\alpha \to \infty$, corresponds to the phase diagram reported in Fig.~\ref{Fig1}(b).
We find that the results obtained for $\alpha \to \infty$ remain qualitatively valid down to $\alpha=1$.
The butterfly structure is essentially preserved for $\mu/J \geq 0$, while it is slightly distorted for negative values of $\mu/J$.
For $\alpha>1$, the QFI is still maximized by considering the operator $\hat{J}_x$, and we identify an xSR phase when $\beta_x=2$ [yellow region in Fig.~\ref{Fig2}(a)].

%%%%%%%%%%%%%%%%%%%%%%%%%%%%%%%%%%
%% FIGURE 3
%%%%%%%%%%%%%%%%%%%%%%%%%%%%%%%%%%
\begin{figure*}[t!]
\begin{center}
\includegraphics[width=1\textwidth]{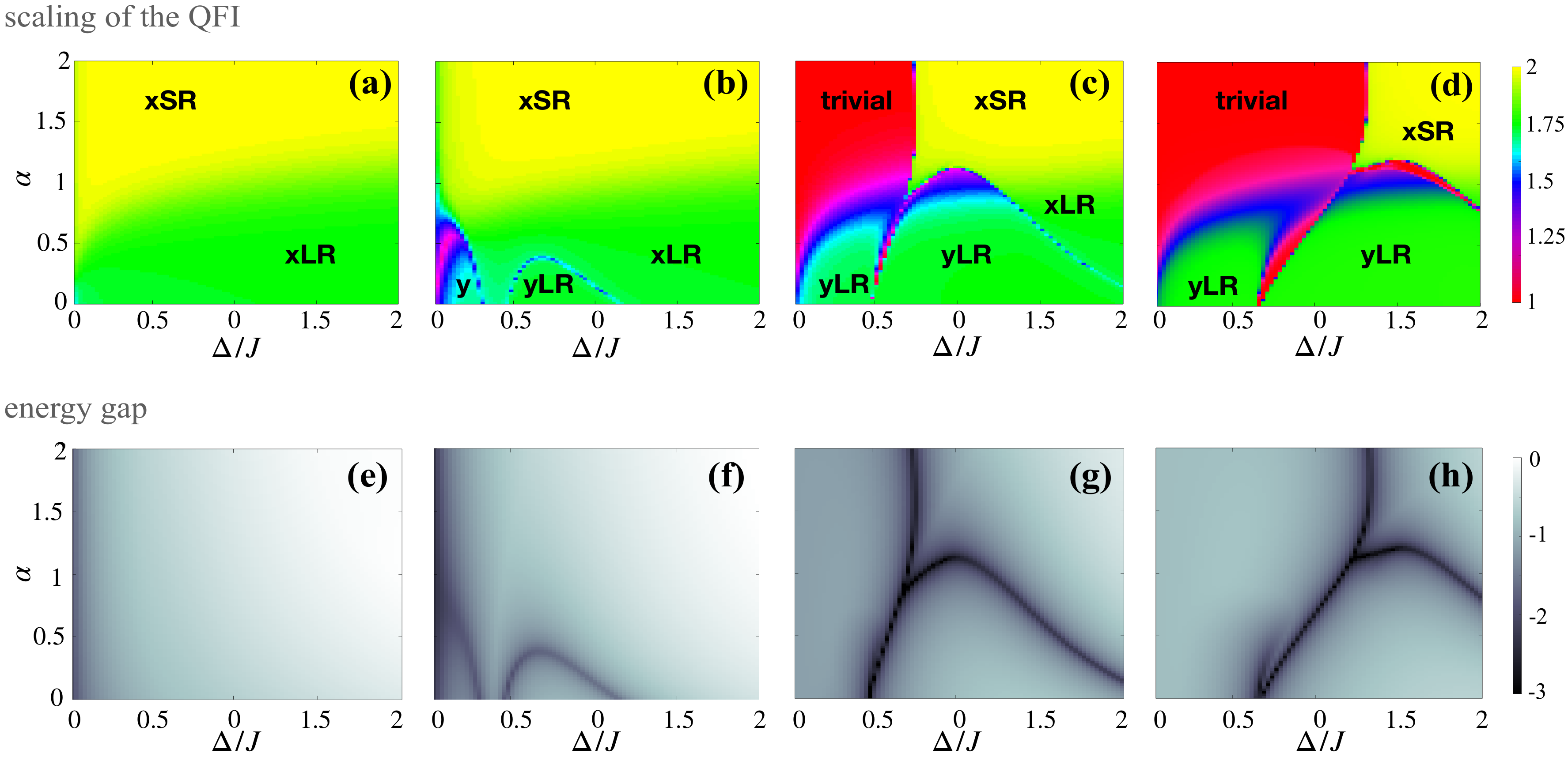}
\end{center}
\caption{Phase diagram of the model in Eq.~(\ref{H}) with commensurate chemical potential, as a function of $\Delta / J$ and $\alpha$. 
Top panels: scaling exponent $\beta$, Eq.~(\ref{eqbeta}). 
Bottom panels: mass gap, $\log_{10}(\delta E/J)$ (with same cutoff as in Fig.~\ref{Fig2}).
Different panels correspond to different values of $V/J$:
$V/J=0.5$~(a,e); $V/J=1$~(b,f); $V/J=1.5$~(c,g) and $V/J=2$~(d,h). 
Here, $\mu/J=0$, $\phi=\pi/2$ and $\omega = 72/117$. 
The scaling of the QFI is calculated between $L=200$ and $L=400$.
}  
\label{Fig3}
\end{figure*}
%%%%%%%%%%%%%%%%%%%%%%%%%%%%%%%%%%
%%%%%%%%%%%%%%%%%%%%%%%%%%%%%%%%%%
%%%%%%%%%%%%%%%%%%%%%%%%%%%%%%%%%%

Probing the predictions of the QFI against a topological invariant is hindered by the difficulty of calculating the latter quantity for finite $\alpha$ in the presence of chemical-potential inhomogeneities (see the Appendix B).
Instead, we have numerically verified that, in an open chain, the xSR phase is characterized by the presence of Majorana modes localized at the edges of the chain.
More specifically, in Fig.~\ref{Fig2}(b) we plot the edge-localization width (ELW), defined as $\delta \ell = (\ell_{\rm left} + \ell_{\rm right})/2$, where
\be \label{Eq.Loc}
\sum_{l=1}^{\ell_{\rm left}} |\psi_l|^2 = C \, ,
\quad {\rm and} \quad
\sum_{l=L-\ell_{\rm right}+1}^{L} |\psi_l|^2 = C \, .
\ee
Here $\psi_l$ is the normalized discrete wavefunction of the first excited eigenstate, which is quasi-degenerate at finite size with the lowest-energy one in the Majorana phase.
In other words, $\ell_{\rm left}$ and $\ell_{\rm right}$ count the number of lattice sites over which $\psi_j$ is spatially localized, with cumulative probability $C$, starting from the left and right edges, respectively.
In practice, we set $2C=0.9$.
In the presence of Majorana modes, one has $\ell_{\rm left} \approx \ell_{\rm right} \ll L/2$, so that $\delta \ell /L \sim 1/L$, whereas for an extended wavefunction $\ell_{\rm left} \approx \ell_{\rm right} \approx L/2$, so that $\delta \ell /L \approx 1$.
Furthermore, the boundaries of the xSR phase are characterized by the closing of the gap $\delta E/J$, as shown in Fig.~\ref{Fig2}(c).

\subsection*{Finite-$\alpha$ regime: $\alpha<1$}

Let us now turn to the case $\alpha <1$. 
In Fig.~\ref{Fig2}(d) and (e), we plot $\beta_x$ and $\beta_y$, respectively.
In each panel, the green region corresponds to a scaling exponent $\beta_{x,y}=7/4$, while the red region corresponds to $\beta_{x,y}=1$.
As we see, we still recover an asymmetric butterfly-like structure, qualitatively similar, especially for $\mu/J \geq 0$, to that obtained for $\alpha>1$.
By analogy with the uniform case~\cite{Pezze2017}, we identify xLR and yLR phases.
Notably, the green and red regions in Fig.~\ref{Fig2}(d) and (e) complement each other, such that $\beta=7/4$ for all values of $\mu/J$, except at the transition between xLR and yLR regions, where we find $\beta_x=\beta_y=3/2$.
This transition is marked by the closing of the mass gap, as shown in Fig.~\ref{Fig2}(f).
We also observe that the presence of massive Dirac modes is not a characteristic feature of the xLR phase: for instance, for parameters in the green region of Fig.~\ref{Fig2}(d), we observe both localized and extended modes, with no distinctive distribution.
In other words, unlike the Majorana modes for $\alpha>1$, the massive Dirac modes for $\alpha<1$ are not robust against inhomogeneities.

Recently, Ref.~\cite{frax2021} reported the calculation of a real-space winding number (adapted from the BDI class~\cite{prodan2014} and here denoted by $\tilde{\nu}$) for the inhomogeneous model considered in this section.
To allow for a direct comparison with the QFI, we investigate in Fig.~\ref{Fig3} the phase diagram in the $\Delta/J$--$\alpha$ plane for different values of $V/J$ (to be compared with Fig.~5 of Ref.~\cite{frax2021}). 
The top panels show the scaling exponent $\beta$, Eq.~(\ref{eqbeta}), while the bottom panels show the corresponding mass gap.  
The different columns correspond to different values of $V/J$: $V/J=0.5$ in panels~(a,e), $V/J=1$ in panels~(b,f), $V/J=1.5$ in panels~(c,g), and $V/J=2$ in panels~(d,h). 
The parameters considered here, $\mu=0$, $\phi=\pi/2$, and $\omega=72/117$, are analogous to those used in Ref.~\cite{frax2021}. 
For relatively small values of $V/J$ and sufficiently large $\Delta/J$, we observe an xSR phase at relatively large $\alpha$ and an xLR phase at small $\alpha$.
The transition between the two phases occurs without closing the energy gap, see Fig.~\ref{Fig3}(e), as allowed by the LR coupling. The same transition is connected with that observed in \cite{vodola2014,vodola2016}.
The phase diagram becomes richer when increasing $V/J$. 
A nested structure of lobes, each hosting both a xLR and a yLR phase, appears at sufficiently small values of $\alpha$ and $\Delta/J$. 
The lobes are separated by massless lines.
Furthermore, the lobe observed at small $\Delta/J$ is again not separated by the trivial phases by any closing the mass gap.
The blue and violet regions in panels~(b), (c), and (d) appear extended and blurred, likely due to finite-size effects.
Overall, our results agree qualitatively well with those of Ref.~\cite{frax2021}: the xSR region corresponds to $\tilde{\nu}=1$, while the xLR and yLR phases correspond to $\tilde{\nu}=0.5$ and $\tilde{\nu}=-0.5$, respectively.
However, the study of the QFI also accesses regions of the phase diagram where the calculation of $\tilde{\nu}$ in Ref.~\cite{frax2021} was hindered by the smallness of the mass gap.
This indicates a richer phase structure than that reported in Ref.~\cite{frax2021}$,$ and promotes the QFI as a valuable tool for investigating the phase diagram of topological systems, beyond the cumbersome analysis of topological invariants.

We finally note that the lobe structure shown in Fig.~\ref{Fig3} can be linked to the butterfly-like structure observed in Figs.~\ref{Fig1} and~\ref{Fig2}, as well as to the fact that, also for $\alpha<2$, only two phases are possible, neither of which breaks the $Z_2$ parity or charge-conjugation symmetry, either explicitly or spontaneously.
Furthermore, for large values of $\Delta/J$, the phase diagram recovers the same qualitative structure as in the limit $V \to 0$, where it is dominated by the topological Majorana phase and by the xLR phases, as expected.
Increasing $\Delta/J$ favors topology for all values of $\alpha$, whereas increasing $V/J$ suppresses it.
This trend can be understood qualitatively in terms of the Hofstadter mapping discussed above~\cite{hofstadter1976}: increasing $V/J$, which is mapped onto the relative hopping amplitude along the second direction (orthogonal to $\hat{j}$), tends to weaken the effect of the pairing $\Delta$, thereby destroying topology for all values of $\alpha$.

\section{Incommensurate quasiperiodic modulation (Aubry-André potential)}

In this section, we analyze the model in Eq.~(\ref{H}) in the presence of onsite disorder.
In particular, we consider the non-uniform chemical potential in Eq.~(\ref{muj_AA}), where $\omega$ is now an {\it irrational} number, here chosen as the inverse golden ratio,
\be \label{Eq.gr}
\omega = \frac{\sqrt{5}-1}{2} \, .
\ee
For $\Delta = 0$, Eq.~(\ref{H}) reduces to the celebrated Aubry--Andr\'e model~\cite{AubriAndre,kraus2012}. 
This model exhibits a metal--insulator (localization--delocalization) transition at the critical value $V_c/J = 1$, which can be determined exactly via a self-duality mapping~\cite{AubriAndre}.
For $V<V_c$, all eigenstates are extended, whereas for $V>V_c$ they are all localized. 
This transition has been observed experimentally using a non-interacting Bose--Einstein condensate trapped in a bichromatic lattice potential~\cite{roati2008}; see also Ref.~\cite{SchreiberSCIENCE2015}, as well as with photons in optical waveguides~\cite{LahiniPRL2009, SegevNATPHYS2013}.  
In the following, we set $\phi=\pi/2$ in Eq.~(\ref{muj_AA}) and study the system with open boundary conditions.

%%%%%%%%%%%%%%%%%%%%%%%%%%%%%%%%%%
%% FIGURE 4
%%%%%%%%%%%%%%%%%%%%%%%%%%%%%%%%%%
\begin{figure}[!]
\begin{center}
\includegraphics[clip,width=\columnwidth]{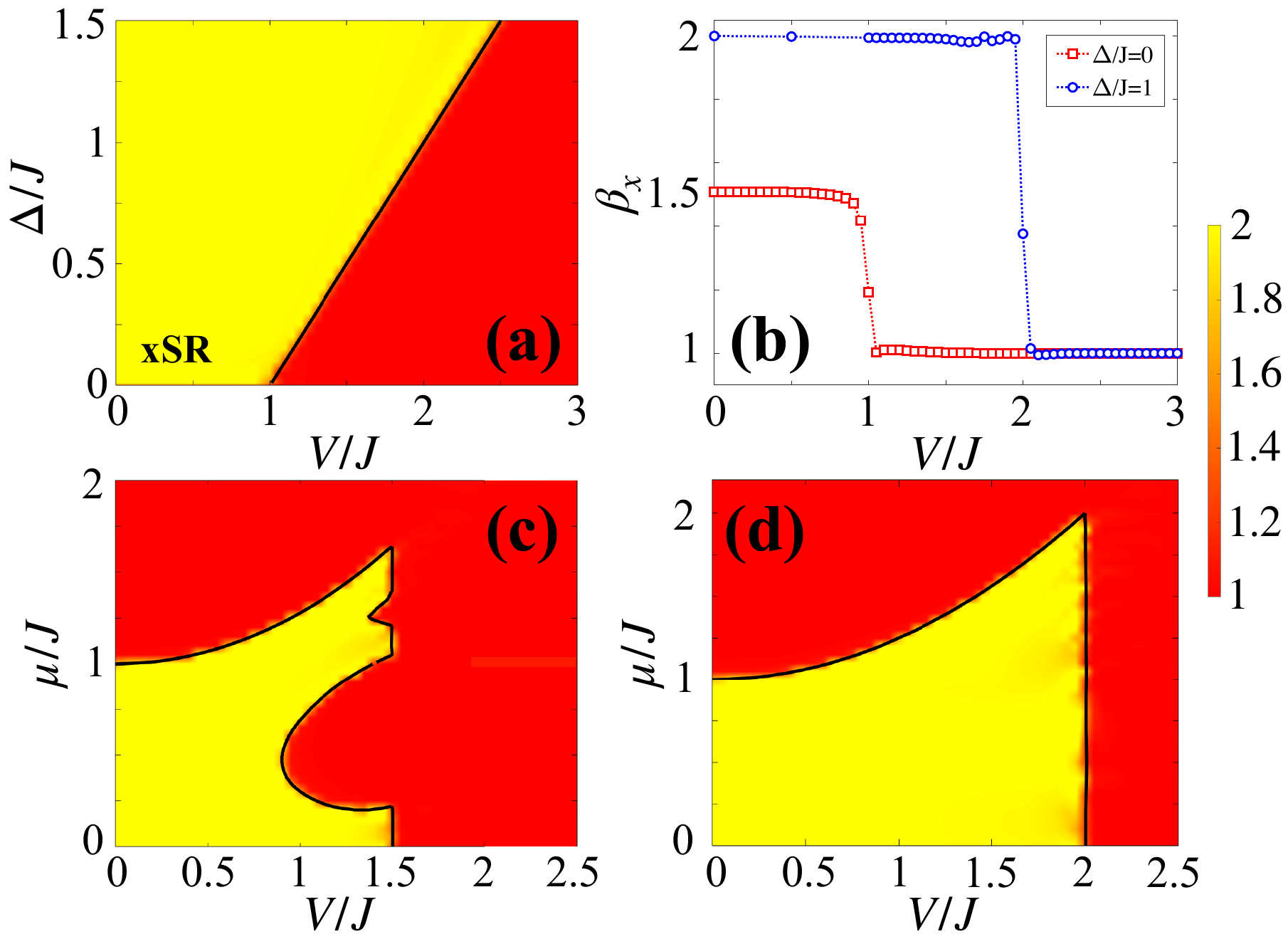}
\end{center}
\caption{
Phase diagram of the model in Eq.~(\ref{H}) with SR couplings ($\alpha \to \infty$) and an incommensurate modulation frequency given by Eq.~(\ref{Eq.gr}).
Panel~(a) shows the scaling exponent $\beta_x$ of the QFI in the $V$--$\Delta$ plane at $\mu/J = 0$.
The black line is Eq.~(\ref{Vcbic}).
Panel~(b) shows cuts through the phase diagram in panel~(a) for $\Delta/J=0$ (red circles) and $\Delta/J=1$ (blue circles).
Panels~(c) and~(d) report the scaling exponent $\beta_x$ in the $\mu$--$V$ plane for $\Delta/J = 0.5$ and $\Delta/J = 1$, respectively.
The black line indicates the transition between a topological phase ($\nu=1$) and a trivial one ($\nu=0$), where $\nu$ is the $Z_2$ topological invariant calculated using Eq.~(\ref{calcnu}) for $L=200$.
Here, the QFI scaling is calculated between $L=100$ and $L=200$ with open boundary conditions.
}
\label{Fig4}
\end{figure}
%%%%%%%%%%%%%%%%%%%%%%%%%%%%%%%%%%
%%%%%%%%%%%%%%%%%%%%%%%%%%%%%%%%%%
%%%%%%%%%%%%%%%%%%%%%%%%%%%%%%%%%%

\subsection*{Limit $\alpha \to \infty$}

For $\Delta \neq 0$, the model in Eq. (\ref{muj_AA}) with $\omega$ given in Eq.~(\ref{Eq.gr}) has been studied in Refs.~\cite{DeGottardiPRL2013, CaiPRL2013, CaiJP2014, TezukaPRB2012} in the limit $\alpha \to \infty$. 
For $\mu=0$, 
a quantum phase transition takes place at a critical value of the disorder strength \cite{DeGottardiPRL2013, CaiPRL2013}
\be \label{Vcbic}
\frac{\vert V_c \vert}{J} = 1 + \frac{\Delta}{J} \, .
\ee
When $V<V_c$, the $Z_2$ topological invariant in Fig. (\ref{calcnu}) is $\nu=1$ and the system thus  hosts Majorana modes. 
Instead, for $V>V_c$, $\nu=0$ holds and the system is in the topologically trivial superconductive phase.

In Fig.~\ref{Fig4} we report the scaling exponent $\beta_x$.
Figure~\ref{Fig4}(a) shows $\beta_x$ in the $V$--$\Delta$ plane for $\mu=0$.
The QFI scaling changes abruptly from superextensive, $\beta_x=2$, in the topological region ($V<V_c$), to extensive, $\beta_x=1$, in the trivial superconducting region ($V>V_c$).
This sharp change occurs exactly along the critical line given in Eq.~(\ref{Vcbic}).
Figure~\ref{Fig4}(b) shows cuts through the phase diagram in panel~(a) for $\Delta/J=0$ (red squares) and $\Delta/J=1$ (blue circles). 
The metallic phase at $\Delta/J=0$ and $V/J=0$ is characterized by $\beta_x=3/2$, which is intermediate between the topological Majorana phase and the trivial one. 
Moreover, the Aubry--Andr\'e localization transition, again at $\Delta/J=0$, is signaled by an abrupt change in the QFI scaling, with $\beta_x$ jumping to $1$ at $V/J=1$.
This sharp transition agrees with the critical value $V_c/J$ predicted by Eq.~(\ref{Vcbic})~\cite{AubriAndre}.

We further study the case $\mu \neq 0$.
The lower panels of Fig.~\ref{Fig4} show the phase diagram of the scaling exponent $\beta_x$ in the $V$--$\mu$ plane, for $\Delta/J = 0.5$ in panel~(c) and $\Delta/J = 1$ in panel~(d).
We note that the phase diagrams in Fig.~\ref{Fig4}(c) and~(d) are mirror-symmetric under $\mu \to -\mu$.
For $|\mu|/J \leq 1$, disorder competes with the Majorana phase, which disappears for sufficiently large values of $V$.
The critical value of $V$ signaling the quantum phase transition is essentially the same as for $\mu/J=0$ when $\Delta/J \geq 1$.
For $\Delta/J<1$ characteristic re-entrances appear, as already noted in Ref.~\cite{CaiJP2014}, and confirmed here by the study of the QFI.  
For $|\mu|/J>1$, disorder may even enlarge the topological phase: for instance, in Fig.~\ref{Fig4}(d), at $\mu/J=1.5$, the phase is topologically trivial for $0 \leq V/J \lesssim 1.5$, while it becomes nontrivial for sufficiently strong disorder, namely for $1.5 \lesssim V/J < 2$. 
The disorder-induced quantum phase transition predicted by the QFI is found to be in perfect agreement with that obtained from the calculation of the $Z_2$ topological invariant~\cite{DeGottardiPRL2013,CaiPRL2013,CaiJP2014}, shown by the black line in Figs.~\ref{Fig4}(c) and~(d).
  
%%%%%%%%%%%%%%%%%%%%%%%%%%%%%%%%%%
%% FIGURE 5
%%%%%%%%%%%%%%%%%%%%%%%%%%%%%%%%%%
\begin{figure}[th!]
\begin{center}
\includegraphics[clip,width=\columnwidth]{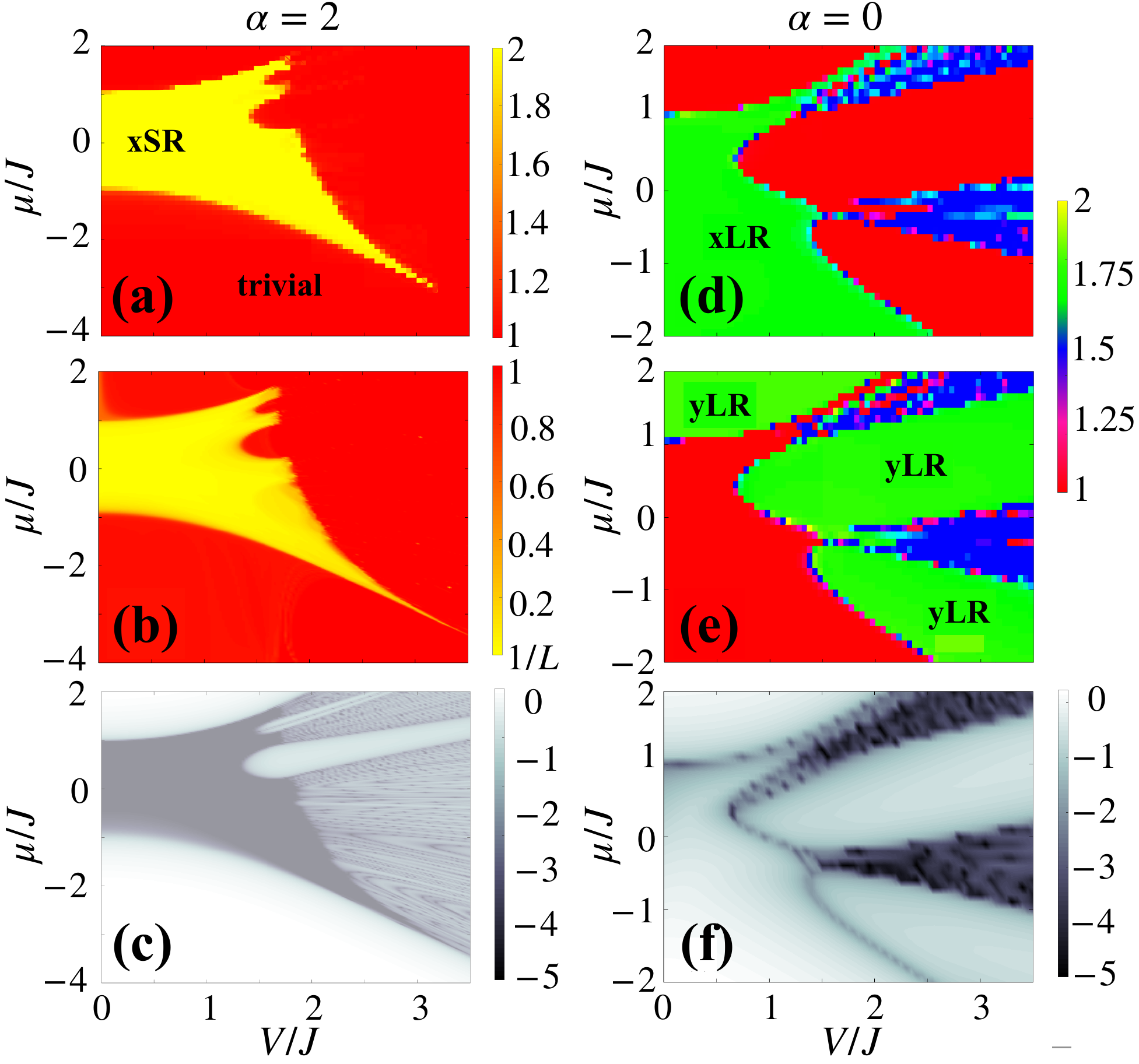}
\end{center}
\caption{
Phase diagram of the model in Eq.~(\ref{H}) with an incommensurate chemical-potential modulation frequency given by Eq.~(\ref{Eq.gr}) and power-law couplings of finite decay exponent ($\alpha<\infty$).
Left panels refer to the case $\alpha=2$ and report, respectively, the scaling exponent $\beta_x$ (a), the edge localization width (b), and the mass gap (c) in the $\mu$--$V/J$ plane for $\alpha=2$.
Right panels instead corresponds to the case $\alpha=0$ and report $\beta_x$ (d), $\beta_y$ (e), and the mass gap (f).
The QFI scaling exponent is calculated between $L=100$ and $L=200$.
The mass gap and the ELW [Eq.~(\ref{Eq.Loc})] are shown for $L=200$.
In all panels, $\Delta/J=1$.
}
\label{Fig5}
\end{figure}
%%%%%%%%%%%%%%%%%%%%%%%%%%%%%%%%%%
%%%%%%%%%%%%%%%%%%%%%%%%%%%%%%%%%%
%%%%%%%%%%%%%%%%%%%%%%%%%%%%%%%%%%

\subsection*{Finite-$\alpha$ regime}

Figure~\ref{Fig5} clarifies how the $\mu$--$V$ phase diagram changes with $\alpha$.
Here $\Delta/J=1$; therefore, the corresponding diagram in the limit $\alpha \to \infty$ is the one shown in Fig.~\ref{Fig4}(d). 
In Fig.~\ref{Fig5}(a) we report the case $\alpha=2$. 
Although the diagram still shows the characteristic transition from $\beta_x=2$ to $\beta_x=1$ (in the trivial superconducting phase), we find qualitative differences with respect to the SR case.
First, the diagram is no longer symmetric with respect to the change of sign of $\mu$. 
Furthermore, for $\mu>0$, characteristic re-entrances appear again, similarly to what was discussed previously.
We further compare the predictions of the QFI with the ELW, calculated from Eq.~(\ref{Eq.Loc}) [see Fig.~\ref{Fig5}(b)], and with the mass gap [see Fig.~\ref{Fig5}(c)].
Overall, we find perfect agreement between the xSR region and the region hosting Majorana edge modes.
As shown in Fig.~\ref{Fig5}(c), the characteristic structures of the phase diagram extend, for $V$ larger than a critical value, into regions where the mass gap is small but finite. 
This feature is supported by finite-size scaling, which suggests saturation as a function of system size.
By contrast, the characteristic reentrances extend into regions with a much larger mass gap.
In Ref.~\cite{CaiJP2014} (for the case $\alpha \to \infty$), the latter feature was associated with a band-gap region, while the former was associated with an Anderson-insulator regime.
The scaling of the QFI does not reveal the difference between these two kinds of regions, since both correspond to a trivial topological phase.

For $\alpha<1$, the phase diagram changes abruptly.
In Fig.~\ref{Fig5}(d) and~(e), we plot $\beta_x$ and $\beta_y$, respectively, for $\alpha=0$. 
The xLR and yLR regions, characterized by $\beta_x=7/4$ and $\beta_y=7/4$, respectively, complement each other, except in extended regions where $\beta_x=\beta_y=3/2$ (shown in blue in the figures), which appear associated with a small mass gap; see Fig.~\ref{Fig5}(f). 
In these regions, significant finite-size effects prevent a precise determination of the phase boundaries, as well as of the corresponding values of the energy gap. 
However, the blue regions and those with low energy gap (shown in darker shades) coincide to satisfactory accuracy.
Finally, no localized Dirac modes are observed for parameter values in this region.

\section{Uncorrelated random disorder (Anderson potential)}

In this section, we consider genuine random disorder {\it \`a la Anderson}~\cite{anderson}, namely
\be \label{anderson}
\mu_l = \mu + \tilde{\mu}_l \, ,
\ee
where $\tilde{\mu}_l$ is a random variable uniformly distributed in the interval $[-V,V]$.
Again, we assume closed (antiperiodic) boundary conditions.
For $\Delta/J = 0$, an arbitrarily small amount of disorder is sufficient to induce localization in one dimension, provided the system is sufficiently large~\cite{LagendijkPT2009, AbrahamsBOOK}.

\subsection*{Limit $\alpha \to \infty$}

This case is qualitatively similar to the incommensurate quasiperiodic case discussed in the previous section: by varying $V/J$, we observe a disorder-induced quantum phase transition from a Majorana phase, characterized by $\bar{\nu} \approx 1$, to a trivial superconducting phase, where $\bar{\nu} \approx 0$. 
Here, $\bar{\nu}$ denotes the disorder average of the $Z_2$ topological invariant defined in Eq.~(\ref{calcnu}) over independent realizations of the disorder.
This disorder-induced quantum phase transition has been characterized by probing the disappearance of the Majorana modes~\cite{DeGottardiPRL2013,CaiPRL2013,hu2015}, as well as by studying the decay of the entanglement entropy and the degeneracies of the entanglement spectrum~\cite{LevyUNIVERSE2019}. 
By contrast, Ref.~\cite{GergsPRB2016} studied this system by examining the presence of long-range order in the correlation functions of the operator $\sigma_x^{(l)}$.

%%%%%%%%%%%%%%%%%%%%%%%%%%%%%%%%%%
%% FIGURE 6
%%%%%%%%%%%%%%%%%%%%%%%%%%%%%%%%%%
\begin{figure}[h!]
\begin{center}
\includegraphics[clip,width=\columnwidth]{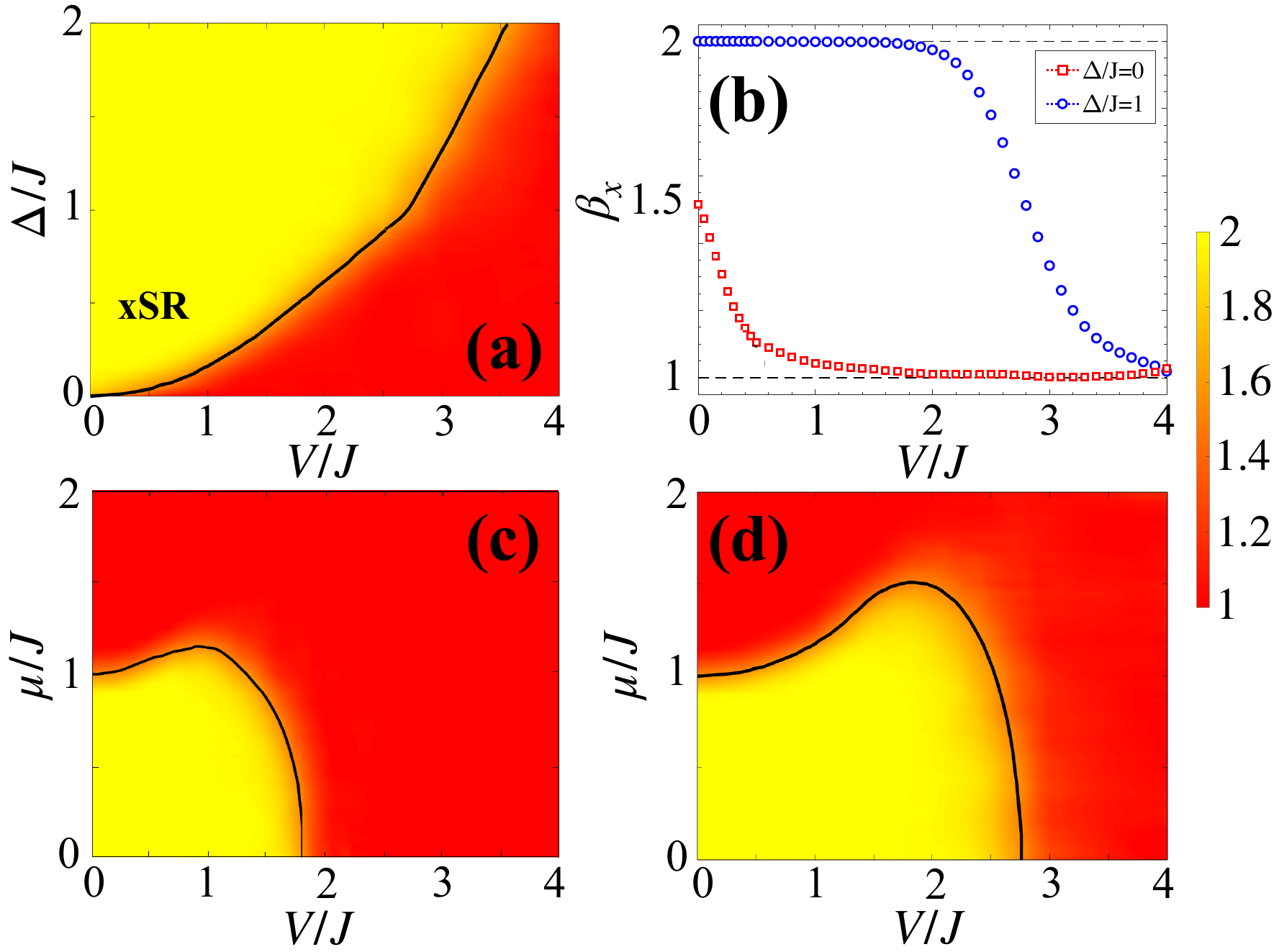}
\end{center}
\caption{
Phase diagram of the model in Eq.~(\ref{H}) with SR couplings ($\alpha \to \infty$) and Anderson disorder, see Eq.~(\ref{anderson}).
Panel~(a) shows $\beta_x$ in the $V$--$\Delta$ plane at $\mu/J = 0$.
The black line corresponds to $\bar{\nu}=0.5$, where $\bar{\nu}$ is the disorder-averaged topological invariant calculated from Eq.~\eqref{calcnu}.
Panel~(b) shows cuts through the phase diagram in panel~(a) for $\Delta/J=0$ (red squares) and $\Delta/J=1$ (blue circles).
Panels~(c) and~(d) show $\beta_x$ in the $V$--$\mu$ plane for $\Delta/J=0.5$ and $\Delta/J=1$, respectively. 
The black line again corresponds to $\bar{\nu}=0.5$.
The obtained phase diagram is symmetric under $\mu \to -\mu$.
Here, we show only the positive $V$ and $\mu$ quadrant.
The QFI scaling exponent is calculated between $L=100$ and $L=200$ with open boundary conditions, while the topological invariant is calculated at $L=200$.
Averaging is performed over 500 disorder realizations.
}
\label{Fig6}
\end{figure}
%%%%%%%%%%%%%%%%%%%%%%%%%%%%%%%%%%
%%%%%%%%%%%%%%%%%%%%%%%%%%%%%%%%%%
%%%%%%%%%%%%%%%%%%%%%%%%%%%%%%%%%%

Here, we calculate the QFI of the ground state of Eq.~(\ref{H}) for several independent realizations of the disordered potential in Eq.~(\ref{anderson}).
We then compute the disorder-averaged QFI and extract the scaling exponent by varying the system size.
Figure~\ref{Fig6} reports the corresponding results.
In panel~(a), we show the scaling exponent $\beta_x$ in the $V$--$\Delta$ phase diagram for $\mu/J=0$.
Similarly to Fig.~\ref{Fig4}, we observe a transition from a regime in which the scaling $\beta_x=2$ is unaffected by disorder to a region at larger $V$, where $\beta_x=1$. 
The black line in the figure indicates the values of $\Delta/J$ and $V/J$ for which $\bar{\nu}=0.5$~\cite{DeGottardiPRL2013}; it sharply separates a topological phase with $\bar{\nu}=1$ (for small $V/J$) from a trivial phase with $\bar{\nu}=0$.
In particular, at $\Delta/J=1$, the quantum phase transition occurs at $V/J=\exp(1)$~\cite{DeGottardiPRL2013}. 
In Fig.~\ref{Fig6}(b), we show $\beta_x$ as a function of $V/J$ for $\Delta/J=0$ (red squares) and $\Delta/J=1$ (blue circles). 
For $\Delta/J=0$, the metallic phase is not robust against disorder, and the scaling exponent $\beta$ decreases rapidly from $\beta_x=3/2$ (at $V=0$) to $\beta_x=1$.
More generally, for $\Delta\neq 0$, the transition between $\beta_x=2$ and $\beta_x=1$ is much smoother than the one observed in Fig.~\ref{Fig4}, due to disorder fluctuations.

Finally, in Fig.~\ref{Fig6}(c) and~(d), we plot the exponent $\beta_x$ in the $V$--$\mu$ plane for $\Delta/J=0.5$ in panel~(c) and $\Delta/J=1$ in panel~(d).
Again, the transition from the Majorana phase to the trivial superconducting phase, along the black line identified by $\bar{\nu}=0.5$, is marked by a change in the scaling exponent $\beta_x$. 
The phase diagram is qualitatively similar to that obtained for the incommensurate Aubry--Andr\'e potential in Fig.~\ref{Fig4}. 
As in the case of the Aubry--Andr\'e potential, disorder enlarges the Majorana phase~\cite{GergsPRB2016,LevyUNIVERSE2019}. 
This effect can be understood from the fact that the disorder term in Eq.~\eqref{anderson} makes $\mu_l$ symmetrically distributed around $\mu$. 
Therefore, starting for instance from a nontopological phase close to the phase transition and for sufficiently small $V$, disorder selects some Hamiltonian configurations, over which we average, that lie in the topological phase~\cite{giuliano2017,lepori2022}, thereby favoring it.
Compared to Fig.~\ref{Fig4}, the disorder-driven transitions in Fig.~\ref{Fig6} are less sharp. 
In fact, in Fig.~\ref{Fig4}, the QFI is evaluated, at fixed $\Delta/J$ and $V/J$, for a single configuration, characterized by a single set $\{\mu_l\}$ in Eq.~\eqref{Eq.gr}. 
On the contrary, in Fig.~\ref{Fig6} and again at fixed $\Delta/J$ and $V/J$, the QFI is averaged on a finite number of configurations, characterized by different values for $\{\mu_l\}$, distributed as in Eq. \eqref{anderson}. 
As a consequence, the resulting disorder-driven transitions are less sharp than in the former case. 
However, this statistical noise can be reduced by increasing the number of configurations over which the QFI is averaged.

%%%%%%%%%%%%%%%%%%%%%%%%%%%%%%%%%%
%% FIGURE 7
%%%%%%%%%%%%%%%%%%%%%%%%%%%%%%%%%%
\begin{figure*}[t!]
\begin{center}
\includegraphics[width=0.9\textwidth]{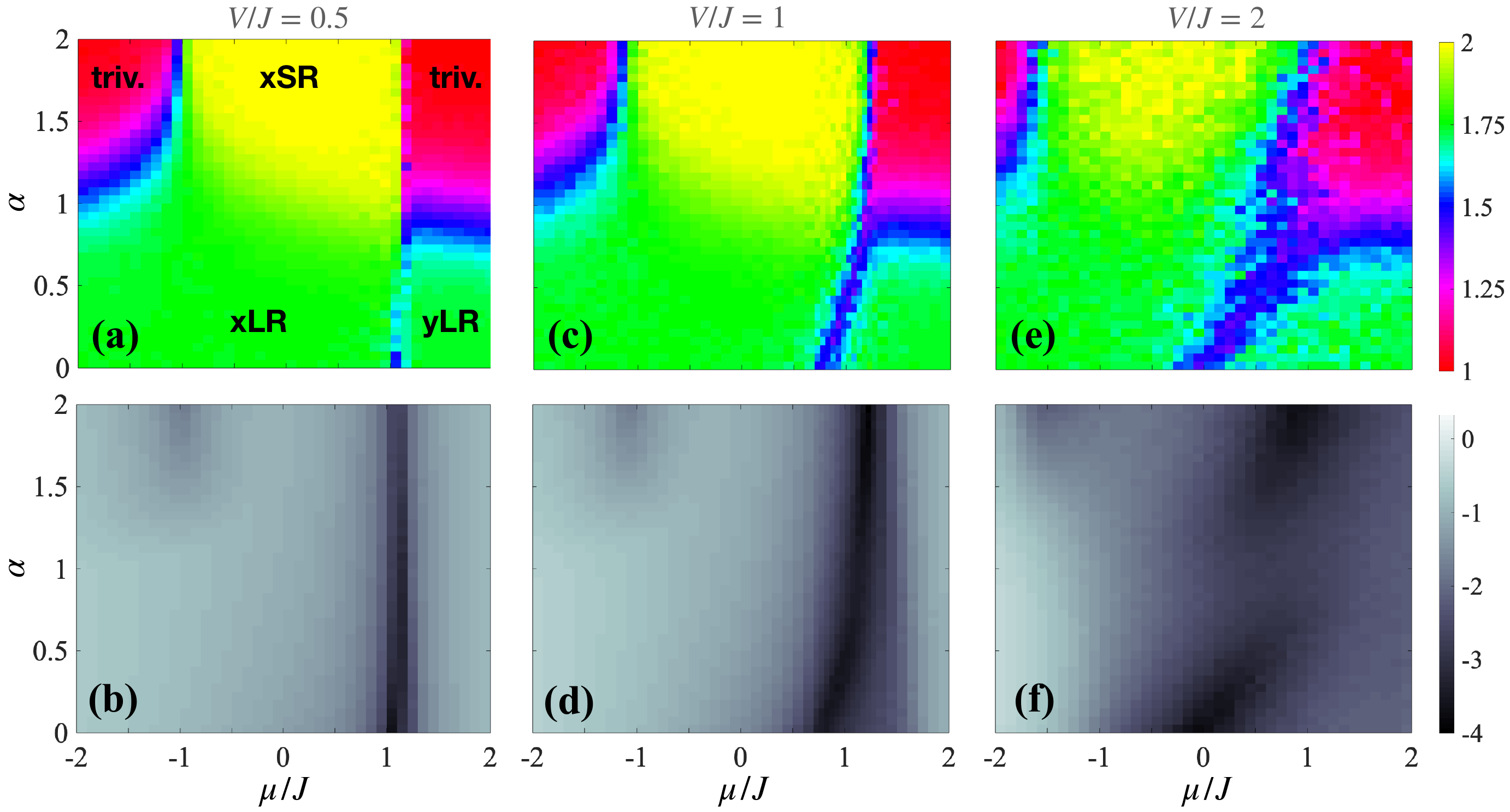}
\end{center}
\caption{
Phase diagram of the model in Eq.~\eqref{H} with Anderson disorder, Eq.~\eqref{anderson}. 
The upper-row panels show \(\beta\) as a function of \(\mu/J\) and \(\alpha\), for different values of \(V/J\): \(0.5\) (a), \(1\) (b), and \(2\) (c). 
The lower-row panels show the corresponding disorder-averaged mass gap, $\log_{10}(\delta E/J)$. 
Here \(\Delta/J=1\), and the other parameters are as in Fig.~\ref{Fig6}. 
In panels (c)--(f), the separation lines become much less well defined: averaging over a larger number of disorder realizations would be required to determine them more accurately.
}
\label{Fig7}
\end{figure*}

\subsection*{Finite-$\alpha$ regime}

For Anderson disorder, the phase diagram in the $V$--$\mu$ plane is very similar to that found in the incommensurate case shown in Fig.~\ref{Fig5}, and we therefore do not report it explicitly here.  
Instead, in the various panels of Fig.~\ref{Fig7}, we plot $\beta$ in the $\mu/J$--$\alpha$ phase diagram for $\Delta/J=1$ and for different values of $V/J$: $0.5$ in panel~(a), $1$ in panel~(b), and $2$ in panel~(c). 
The corresponding diagram for $V=0$ (namely, for the homogeneous case $\mu_l=\mu$) was studied in Ref.~\cite{pezze2017}.
For sufficiently small values of $V$ [e.g., $V/J=0.5$ in panel~(a)], the phase diagram shows features very similar to those found at $V=0$~\cite{vodola2014}. 
In particular, we recognize an xSR phase for $\alpha>1$ and $|\mu|/J\leq 1$, an xLR phase for $\mu/J \lesssim 1$ and $\alpha<1$, where $\beta=\beta_x=7/4$ -- again with the separation not occurring exactly at $\alpha=1$ because of finite-size effects, as already inferred in Refs.~\cite{vodola2014,vodola2016} -- and a yLR phase for $\mu/J \gtrsim 1$ and $\alpha<1$, where $\beta_y=7/4$. 
The transition between the xLR and yLR phases at $\alpha<1$, as well as that between the xSR and trivial phases at $\alpha>1$, is marked by a closing of the mass gap around $\mu/J=1$, where $\beta_x=\beta_y=3/2$. 
Upon increasing $V$, the most relevant effect is the enlargement of the yLR phase, which closely follows the behavior of the mass gap; see Fig.~\ref{Fig7}(b), (d), and (f). However, all separation lines become much less sharp.

\section{Conclusions}

To summarize, in this work we have discussed the stability of multipartite entanglement (ME) against inhomogeneities in paradigmatic gapped fermionic wires, which may also host symmetry-protected topological phases.
In particular, we have calculated the QFI of the ground state of a generalization of the celebrated Kitaev chain with variable-range pairing, in the presence of periodic, quasiperiodic, and genuinely random offsets. 
Our analysis conveys two main  messages.

First, the QFI scaling exponent allows one to identify clearly the topological and long-range regions, at least in the notable cases considered here. 
By contrast, topological invariants are often difficult to define and/or compute in the presence of inhomogeneities and long-range pairing.  
In particular, in the limit $\alpha \to \infty$, the topological regions identified by the QFI agree perfectly with those singled out by the $Z_2$ topological invariant.
For $\alpha<\infty$, the boundaries between the different phases agree well with the behavior of the mass gap whenever the latter is relevant for determining a transition between trivial and topological phases.
Second, the ME identified by the QFI appears robust against inhomogeneities.
In some cases, the QFI is even enhanced by disorder.
In particular, the superextensive scaling of the QFI is as robust as the corresponding nontrivial phases themselves: the stability of ME persists as long as the SR and LR topological phases are protected by the Hamiltonian symmetries and by a finite mass gap, neither of which is destroyed by the added inhomogeneities.
Overall, this work promotes the QFI as a useful tool to analyze  topological systems, even in the presence of inhomogeneities and disorder.  \\

\section*{Acknowledgements}
We thank M. Burrello,  A. Nava, and A. Smerzi for discussions.
We acknowledge financial support from the European Union’s Horizon 2020 research and innovation program - Qombs Project, FET Flagship on Quantum Technologies grant no. 820419.
L. L. also acknowledges research funding from the PRIN project 2017 number 20177SL7HC, financed by the Italian Ministry of education and research.
L. L. also acknowledges financial support by a project funded under the National Recovery and Resilience Plan (NRRP), Mission 4 Component 2 Investment 1.3 - Call for tender No. 341 of 15/03/2022 of Italian Ministry of University and Research funded by the European Union NextGenerationEU, award number PE0000023, Concession Decree No. 1564 of 11/10/2022 adopted by the Italian Ministry of University and Research, CUP D93C22000940001, Project title ”National Quantum Science and Technology Institute” (NQSTI), spoke 2.

\newpage
\onecolumngrid

\section*{APPENDIX A: derivation of the bound in Eq. \eqref{boundQFI} of the main text for pure quantum states.}
\label{proofbound}

We demonstrate that Eq.~(\ref{Fxyabs}) i the main text
%, 
%\be \label{proofA1}
%\tilde F_Q\big[|\psi\rangle,\hat J_\alpha\big]
%=
%\sum_{l,m=1}^{L}\big|\rho_{\alpha}^{(l,m)}\big| \, ,
%\qquad \alpha=x,y,
%\ee
%with 
%$\rho_{\alpha}^{(l,m)}
%\equiv
%4\,\langle \psi|\hat o_l^{(\alpha)}\hat o_m^{(\alpha)}|\psi\rangle_c$,
has bounds depending on $k$-partite entanglement in the generic pure state $\ket{\psi}$.
Assume that $|\psi\rangle$ is $k$-producible, namely
$|\psi\rangle=\bigotimes_{a=1}^{h}|\phi_a\rangle$,
where the subsets $D_a$ contain $n_a\le k$ sites and satisfy $\sum_{a=1}^{h} n_a=L$.
Since connected correlations vanish between different factors of a product state, one has
\be
\langle \psi|\hat o_l^{(\alpha)}\hat o_m^{(\alpha)}|\psi\rangle_c=0 \, ,
\qquad
l\in D_a \, ,\; m\in D_b\, ,\; a\neq b \, .
\ee
Therefore,
\be
\tilde F_Q\big[|\psi\rangle,\hat J_\alpha\big]
=
\sum_{a=1}^{h}\sum_{l,m\in D_a}\big|\rho_{\alpha}^{(l,m)}\big| \, .
\ee
If the local normalization is such that $|\rho_{\alpha}^{(l,m)}|\le 1$ for all $l,m$, then each ordered pair inside a block contributes at most $1$, and thus
$\sum_{l,m\in D_a}\big|\rho_{\alpha}^{(l,m)}\big|
\le n_a^2$.
Hence
\be
\tilde F_Q\big[|\psi\rangle,\hat J_\alpha\big]
\le
\sum_{a=1}^{h} n_a^2 \, .
\ee
Writing $L=sk+r$, for $0\le r<k$, the maximum of $\sum_a n_a^2$ under the constraints $n_a\le k$ and $\sum_a n_a=L$
is obtained for $s$ blocks of size $k$ and one block of size $r$, which gives
\be
\tilde F_Q\big[|\psi\rangle,\hat J_\alpha\big]
\le
sk^2+r^2 \le kL \, .
\ee
Therefore, if $\tilde F_Q\big[|\psi\rangle,\hat J_\alpha\big] > sk^2+r^2$, then the state cannot be $k$-producible and must contain at least $(k+1)$-partite entanglement~\cite{HyllusPRA2012,TothPRA2012}.
The above proof follows the one in Ref.~\cite{lepori2021,notaSM2} and -- since Eq.~\eqref{Fxyabs} involves the absolute value of two-point correlations, $\vert \rho_{\alpha}^{(l,m)}\vert $ -- it is valid for pure states. 
Instead, it fails for general $k$-producible mixed state because a convex mixture of product states can carry classical inter-block correlations, and those generate nonzero connected correlators even without entanglement.

The above proof can be straightforwardly extended to Eq.~(\ref{Fxy}), in the case $s_l, s_m=\pm 1$.
For the case of correlators $s_l  s_m \rho_{\alpha}^{(l,m)}$, the $k$-producible bounds also apply to $F_Q[\hat{\rho}, \hat{J}_{x,y}(\vect{s})]$, for a generic mixed state $\hat{\rho}$.

\section*{APPENDIX B: calculation of the topological invariant in Eq. \eqref{berryphase} of the main text.}

In the main text, the QFI is compared with topological indices, since both can identify distinct quantum phases.
For completeness, we briefly review these indices here for finite \(\alpha\), and highlight their main limitations in the cases considered in this work.
For \(1<\alpha<\infty\), the topological index \(\nu\) can be computed either from Eq.~\eqref{berryphase}, assuming antiperiodic boundary conditions, or from the Pfaffian. 
By contrast, the transfer-matrix approach leading to Eq.~\eqref{calcnu} is technically cumbersome.
For \(\alpha<1\), the index \(\nu\) cannot be defined in the same way. 
In this regime, under closed boundary conditions, one may instead consider~\cite{leporiLR}
\beq
\frac{1}{2}+\frac{i}{\pi}\int_{BZ}\mathrm{d}k\,\langle u_k|\partial_k u_k\rangle
=
\frac{1}{2}+\nu \, ,
\label{berryphaseLR}
\eeq
which equals \(1\) for \(\mu/J<1\), where massive edge modes are present, and \(0\) for \(\mu/J>1\), where protected edge states are absent. 
By contrast, Eq.~\eqref{berryphase} becomes ill-defined for \(\alpha<1\), yielding semi-integer values~\cite{leporiLR,frax2021}.
The critical line \(\alpha=1\), while marking a transition beyond first order, remains gapped in the thermodynamic limit. 
As a consequence, the \(Z_2\) parity symmetry is restored even with open boundary conditions, as also reflected in the entanglement spectrum~\cite{leporiLR}. 
These features originate from the algebraic tails of the two-point correlations, which persist even in massive phases. 
Moreover, along the critical line \(\mu/J=1\), conformal invariance breaks down for \(\alpha<1\). 
These properties are closely related to the singular states at the edge of the Brillouin zone, whose energy diverges~\cite{lepori2016eff}.
A useful quantity to distinguish long-range phases, and to separate them from the short-range one, is still the Pfaffian. 
In momentum space, the relevant sign reduces to~\cite{KitaevPU2001}
\beq
\zeta=\mathrm{sign}\!\Big[\mathrm{Pf}\big(\hat M(0)\big)\,\mathrm{Pf}\big(\hat M(\pi)\big)\Big].
\label{pf2}
\eeq
Indeed, charge-conjugation symmetry in class \(D\) implies that gap closings of \(\hat H(k)\), and equivalently of the antisymmetric matrix \(\hat M(k)\), occur in opposite-momentum pairs \(\pm k\), which do not affect the Pfaffian sign. 
Only the self-conjugate points \(k=0,\pi\) can change their sign, and these are precisely the points where the gap closes at the transitions of the long-range Kitaev chain in Eq.~\eqref{H}.
For \(\alpha<1\), however, the Pfaffian is ill-defined at \(k=0\), where \(\hat M(k)\) and \(\hat H(k)\) become singular and satisfy
\[
\lim_{k\to 0^-}\hat M(k)\neq \lim_{k\to 0^+}\hat M(k) \, .
\]
This is the ``second-type singularity'' of Ref.~\cite{leporiLR}, characteristic of long-range phases, in contrast to ``first-type singularities,'' for which the two limits coincide. 
Nevertheless, long-range phases can still be distinguished by the sign of \(\mathrm{Pf}[\hat M(\pi)]\), which takes values \(\tilde\zeta=\pm1\) as the Hamiltonian parameters vary. 
This sign, however, is not related to the fermion-number parity of the ground state, which remains positive because the edge modes are massive. 
Therefore, although it still changes at a second-order transition, it is not a genuine topological invariant, since it is no longer tied to the number of edge modes.
When spatial inhomogeneities are introduced, the Pfaffian sign ceases to be a reliable indicator of short-range or long-range phases. 
This occurs, for instance, in Fig.~\eqref{Fig2}, where the Pfaffian may oscillate rapidly even without gap closings, or fail to change sign when a closing occurs. 
The reason is that, in the presence of inhomogeneities, the Pfaffian can become extremely small in modulus, already for moderate system sizes and in finite regions of parameter space, compared with the characteristic energy scale \(J\). 
Its sign then becomes numerically unstable. 
Moreover, some transitions involve gap closings between degenerate lowest-energy states associated with momenta different from \(0\) and \(\pi\), though still related by charge conjugation~\cite{KitaevPU2001}. 
In such cases, the Pfaffian does not change sign, even when different phases are separated.

\end{document}